\documentclass{appolb}

\begin{document}

\title{Elementary derivation of the Lense-Thirring precession}

\author{Olga Chashchina$^1$, Lorenzo Iorio$^2$ and  Zurab Silagadze$^{1,3}$
\address{$^1$ Novosibirsk State University, 630 090, Novosibirsk, Russia}
\address{$^2$ INFN-Sezione di Pisa, Viale Unit$\grave{\rm a}$ di Italia 68,
70125, Bari, Italy}
\address{$^3$ Budker Institute of Nuclear Physics, 630 090, Novosibirsk, 
Russia}}


\maketitle

\begin{abstract}
An elementary pedagogical derivation of the Lense-Thirring
precession is given based on the use of Hamilton vector. The
Hamilton vector is an extra constant of motion of  the
Kepler/Coulomb problem related simply to the more popular
Runge-Lenz vector. When a velocity-dependent Lorentz-like
gravitomagnetic force is present, the Hamilton vector, as well as
the canonical orbital momentum are no longer conserved and begin
to precess. It is easy to calculate their precession rates, which
are related to the Lense-Thirring precession of the orbit.
\end{abstract}

\section{Introduction}

Every relativistic theory of gravitation must include a Lorentz-like force
induced by a magnetic-type component of the gravitational field; the general
theory of relativity by Einstein does so. More precisely, in its weak-field
and slow-motion approximation the highly nonlinear Einstein field equations
get linearized, thus resembling the linear equations of the Maxwellian
electromagnetism. As a consequence, a magnetic-type component of the
gravitational field appears, induced by the off-diagonal components
$g_{0i}, i=1,2,3$ of the spacetime metric tensor \cite{1-1}. The role of the
electric currents is played by mass-energy currents: in the case of a slowly
rotating mass, a test particle  geodesically moving far from it  is acted upon
by a non-central velocity-dependent Lorentz-like force. For other effects
induced by the gravitomagentic field on the motion of gyroscopes, test
particles, moving clocks and atoms,  light rays see, for example, \cite{1-2}.

Recent years have seen increasing efforts, both from theoretical and
observational points of view, towards a better comprehension of the
post-Newtonian gravitomagnetic field: for a comprehensive recent overview see,
for example, \cite{1-3}. Thus, we feel it is not worthless to offer to the
reader an elementary derivation of the precessional effects induced by the
gravitomagnetic field on some Keplerian orbital elements of a test-particle,
i.e. the so called Lense-Thirring effect \cite{1-4}, although a recent
historical analysis \cite{1-5} suggests that it should be more appropriately
named Einstein-Thirring-Lense effect. For other derivations of such an effect
see, for example, \cite{1-6,1-6P,1-7,1-8,1-9}.

Our presentation is aimed for the first year physics students with limited 
experience in mathematics and theoretical methods of physics. We, therefore,
give somewhat detailed exposition which an experienced physicist may find
unnecessary long, but as our teaching experience shows such an exposition
is helpful and necessary for newcomers in the field. We hope the material 
presented here will help students which just begin their physics education 
not only master some simple mathematical methods learned during the first 
year mechanics course but also feel the beauty of advanced topics which
they will learn in detail later on.

\section{Gravitomagnetism}
Nowadays it is widely accepted that the correct theory of gravitation is
provided by the Einstein's general theory of relativity \cite{2-1} (for an
excellent introduction for beginners see \cite{2-2}; a classic  textbook is
\cite{2-3}). However, the general theory of relativity is a highly non-linear
theory. The spacetime in it is not merely a static background for physical
processes; it is dynamic and affected by any contribution to the
energy-momentum tensor of the system under investigation. Such a tensor enters
in a prescribed manner the Einstein's equations which determine the ten
components of the metric tensor. This metric tensor enters by itself  the
equations of motion of the system under study. There is no hope for a neophyte
to struggle his/her way into this impossibly tangled up mess of non-linear
jungles ``unless months of study on the specialized terminology, procedures,
and conventions of the general relativity theorist have been completed''
\cite{2-4}.

However, virtually the large part of all the performed and/or proposed  tests,
aiming to observationally scrutinize the post-Newtonian regime, deal with
weak fields and non-relativistic velocities.
Therefore, we may expect that the full machinery of general relativity is not
necessarily required to estimate the relevant post-Newtonian effects. Some
linearized version of the theory will suffice to do the job much faster and
physically in a more transparent manner \cite{1-2,2-4,2-5,2-6}.

Although, in general, spacetime is dynamical and there is no natural way to
split it into space plus time, for stationary spacetimes, like the one around
the Earth, stationarity dictates a preferred way of how this splitting can be
performed. There is no approximation in such slicing of spacetime into
three-dimensional space plus one-dimensional time. It is just a new
mathematical language convenient in stationary situations \cite{2-6}.

Under $3+1$ slicing, the spacetime metric tensor $g_{\mu\nu}$ naturally
decomposes into several parts. In the case of weak gravity and
non-relativistic velocities, this decomposition allows one to set up a
remarkable analogy with electromagnetism \cite{1-2,2-4,2-5,2-6}. In particular,
gravitational analogies of the electromagnetic scalar and vector potentials,
$\Phi$ and $\vec{A}$ respectively, are determined by the time-time and
time-space components of the spacetime metric:
$$\Phi=\frac{1}{2}\left (g_{00}-1\right )c^2,\;\;\; A_i=g_{0i}c^2,$$
where $c$ is the speed of light. Gravitoelectric field $\vec{E}$ and
gravitomagnetic field $\vec{H}$ are related to these potentials (in the 
Lorentz gauge \cite{1-2}) in the usual way (up to an extra factor 4 which 
also appears in the Lorentz gauge condition):
$$\vec{E}=-\nabla \Phi-\frac{1}{4c}\frac{\partial \vec{A}}{\partial t} ,
\;\;\; \vec{H}=\nabla\times\vec{A}$$
and satisfy a gravitational analog of Maxwell's equations \cite{1-2,2-6,2-7}
\begin{eqnarray}
&\nabla\cdot \vec{E} =-4\pi G\rho, \qquad &
\nabla\cdot \vec{H} =0 \nonumber \\
&\nabla\times \vec{E} =0,\qquad \qquad &
\nabla\times \vec{H} =4\left [-4\pi G\,\frac{\rho\vec{v}}{c}+
\frac{1}{c}\frac{\partial \vec{E}}{\partial t} \right ].
\label{gme}
\end{eqnarray}
Similarity with  Maxwell's electromagnetic equations is apparent. The role of
charge density is played by mass density, $\rho$, times Newton's
gravitation constant, $G$. The mass current density, $G\rho\vec{v}$, with
$\vec{v}$ as the velocity of the source mass, plays the role of the charge
current density.

However, there are several important differences \cite{2-6}. Gravity is
mediated by a spin-two field and is attractive. In contrast, electromagnetism
is mediated by a spin-one field and can be both attractive and repulsive.
This difference leads to the extra minus signs in the source terms in
(\ref{gme}). Another remnant of the tensor character of gravity is an extra
factor, $4$, in the equation for $\nabla\times \vec{H}$. No gravitational
analog of the Faraday induction is present in (\ref{gme}). However, this is an
artifact of restricting to the only first order terms in $v/c$ of the
gravitating masses in the above equations. At the second order,
$(-1/c)(\partial\vec{H}/\partial t) $ term reappears in the equation for
$\nabla\times \vec{E}$, but also some non-Maxwell-like terms are introduced
elsewhere in the field equations \cite{2-5,2-6}.

Once the gravitoelectric and gravitomagnetic fields are known, the force
acting on a small test body of mass $m$ is given by the formula which is
analogous to the Lorentz force law \cite{2-6}
$$\vec{F}=m\vec{E}+\frac{m}{c}\,\vec{v}\times\vec{H}.$$
More precisely, there is still another difference from electromagnetism as far
as the equation of motion is concerned. So far nothing was said about the
space-space components, $g_{ij}$, of the metric tensor. In general, they do not
correspond to Euclidean space but to that of curved space. Just the space
curvature effects are responsible for the classic general relativistic
predictions for Mercury's perihelion precession and bending of light rays. In
the presence of space curvature, we have the curvilinear equation of motion
\cite{2-4}
\begin{equation}
\frac{1}{\Gamma}\,\frac{d(\Gamma v_i)}{dt}-\frac{1}{2}
\left (\frac{\partial g_{jk}}{\partial x^i}\right )v^jv^k=E_i+
\frac{1}{c}\,(\vec{v}\times\vec{H})_i-\frac{1}{c}\frac{\partial
H_i}{\partial t},
\label{eqm}
\end{equation}
where
$$\Gamma\approx \left (1+\frac{2\Phi}{c^2}-\frac{v^2}{c^2}\right )
^{-1/2} $$
and, as usual, summation over repeated indexes is assumed.

In this paper, we are interested in tiny secular effects of Earth's rotation
on a satellite orbit. Therefore, we will neglect the space curvature effects,
as well as terms of the order of $(v/c)^2$ other than gravitomagnetic.
Although these curvature effects are even more prominent than the effects we
are interested in, they are well known. Therefore, for our goals, equation
(\ref{eqm}) takes the form
\begin{equation}
m\,\frac{d\vec{v}}{dt}=-\frac{\alpha}{r^2}\,\vec{n}+\frac{m}{c}\,\vec{v}
\times \vec{H},
\label{seqm}
\end{equation}
where $\vec{r}=r\vec{n}$ is the radius-vector of the satellite of mass $m$,
$\alpha=GmM$, where $M$ is the Earth's mass, and $\vec{H}$ is the
gravitomagnetic field caused by Earth's rotation.

The gravito-electromagnetic analogy implied by (\ref{gme}) allows us to find
$\vec{H}$ \cite{2-7}. It is well known that the magnetic moment
$$\vec{\mu}=\frac{1}{2c}\int [\vec{r}\times\vec{j}]\,dV,$$
related to the electric current density $\vec{j}$, creates a dipole magnetic
field
$$\vec{H}=\frac{3\vec{n}\,(\vec{n}\cdot\vec{\mu})-\vec{\mu}}{r^3}.$$
Equations (\ref{gme}) indicate that the gravitational analog of the electric
current density is $-4G\rho\vec{v}$. Therefore the gravitational analog of the
magnetic moment is
$$\vec{\mu}=-4G\,\frac{1}{2c}\int \rho\,[\vec{r}\times\vec{v}]\,dV=-2G\,
\frac{\vec{S}}{c},$$
where $\vec{S}=\int [\vec{r}\times\vec{v}]\,\rho dV$ is the rotating body's
proper angular momentum.

Thus the gravitomagnetic field of the Earth is given by
\begin{equation}
\vec{H}=\frac{2G}{c}\,\frac{\vec{S}-3\vec{n}\,(\vec{n}\cdot\vec{S}
)}{r^3}.
\label{egmf}
\end{equation}

\section{Larmor precession}
It is instructive first to consider in detail a simpler case of a
non-relativistic Coulomb atom slightly perturbed by a weak and uniform
constant magnetic field $\vec{H}$. The equation of motion is
\begin{equation}
m\dot{\vec{v}}=-\frac{\alpha}{r^2}\,\vec{n}+\frac{e}{c}\,\vec{v}\times
\vec{H},
\label{larmor}
\end{equation}
where now $m$ and $e$ are electron mass and charge respectively, and $\alpha$
is the strength of the corresponding electromagnetic coupling between the
electron and the atom nucleus.

In the absence of perturbation, $\vec{H}=0$, equation (\ref{larmor}) describes
the electron motion on the ellipse. It is convenient to characterize the
orientation of this ellipse by the angular momentum vector $\vec{L}=\vec{r}
\times\vec{p}$, which is perpendicular to the orbit plane, and the Runge-Lenz
vector
\begin{equation}
\vec{A}=\vec{v}\times\vec{L}-\alpha\,\vec{n},
\label{Runge-Lenz}
\end{equation}
which is directed towards perihelion of the orbit and equals in magnitude to
the orbit eccentricity times electromagnetic coupling $\alpha$. The Runge-Lenz
vector is an extra constant of motion of the Coulomb problem and its existence
is related to the so called hidden symmetry of the problem \cite{3-1,3-2}
which makes the Kepler/Coulomb problem an interesting testing ground for
various algebraic and geometric methods \cite{3-3,3-4,3-5}. Note that first
integrals for the Kepler Problem, equivalents of the Runge-Lenz vector,
were first obtained by Ermanno and Bernoulli many years before Gibbs devised
his vectorial notation \cite{3-6,3-7,3-8}. ``The lack of recognition of
Ermanno and Bernoulli parallels that of Aristarchos and makes one wonder if
there is some sinister influence for those who are initiators in the field of
the motions of the planets'' \cite{3-6}.

The Runge-Lenz vector $\vec{A}$ and the angular momentum vector $\vec{L}$ are
mutually perpendicular. Therefore we can find the third vector $\vec{u}$ such
that
\begin{equation}
\vec{A}=\vec{u}\times\vec{L}.
\label{AuL}
\end{equation}
This vector is called the Hamilton vector \cite{3-9,3-10,3-11,3-12} and it has
the form
\begin{equation}
\vec{u}=\vec{v}-\frac{\alpha}{L}\,\vec{e}_\varphi,
\label{Hamilton}
\end{equation}
where
\begin{equation}
\vec{e}_\varphi=\frac{1}{L}\,\vec{L}\times\vec{n}=\vec{l}\times\vec{n},\;\;\;
\mathrm{with}\;\;\;\vec{l}=\frac{\vec{L}}{L},
\label{e_phi}
\end{equation}
is the unit vector in the direction of the polar angle $\varphi$ in the orbit
plane. Sometimes it is more convenient to characterize the orbit orientation
by vectors $\vec{u}$ and $\vec{L}$.

Now let us return to (\ref{larmor}) with the non-zero magnetic field $\vec{H}$.
As the Lorentz force, in general, has a component outside the unperturbed
orbital plane, it is evident that the motion will no longer be planar. However,
for weak magnetic fields, the Lorentz force component in the orbital plane is
much smaller than the binding Coulomb force. Therefore, it can cause only
small perturbations of the elliptical orbit. The Lorentz force component
perpendicular to the orbital plane is expected to cause this plane to turn
slowly around the Coulomb field center. This intuitive picture of the
perturbed motion suggests us to decompose the electron velocity in such a way
\begin{equation}
\vec{v}=\vec{v}^{\,\prime}+\vec{\Omega}\times\vec{r}.
\label{vvprime}
\end{equation}
Here $\vec{v}^{\,\prime}$ is the electron velocity relative to the
instantaneous orbital plane, and the second term, $\vec{\Omega}\times\vec{r}$,
is due to slow revolution of this plane with a small angular velocity
$\vec{\Omega}$.

If we want the angular momentum $\vec{L}$ to be perpendicular to the instant
orbital plane, we should replace $\vec{v}$ by $\vec{v}^{\,\prime}$ in its
definition:
\begin{equation}
\vec{L}=\vec{r}\times\vec{P},\;\;\;\vec{P}=m\vec{v}^{\,\prime}.
\label{Ldef}
\end{equation}
Analogously, the appropriately generalized Hamilton vector is
\begin{equation}
\vec{u}=\vec{v}^{\,\prime}-\frac{\alpha}{L}\,\vec{e}_\varphi=
\frac{\vec{P}}{m}-\frac{\alpha}{L}\,\vec{e}_\varphi,
\label{Hgen}
\end{equation}
where $\vec{e}_\varphi$ is still given by equation (\ref{e_phi}).

The Runge-Lenz vector, defined as earlier by (\ref{AuL}), is no longer
conserved. However, at instances when the particle is at perihelion of the
instant ellipse, the Runge-Lentz vector still points towards the perihelion,
as the following simple argument shows \cite{3-13}. At the perihelion of the
instant ellipse, the intrinsic velocity $\vec{v}^{\,\prime}$ has no radial
components and, therefore, is perpendicular to the radius vector $\vec{r}$.
Then (\ref{Hgen}) shows that the Hamilton vector is also perpendicular to
$\vec{r}$ and, according to (\ref{AuL}), the Runge-Lenz vector $\vec{A}$ will
be parallel to $\vec{r}$, that is pointing towards the perihelion.

If our intuitive picture of the orbit plane precession is correct, we should
have
\begin{equation}
\dot{\vec{L}}=\vec{\Omega}\times\vec{L}.
\label{Ldot}
\end{equation}
On the other hand,
\begin{equation}
\dot{\vec{L}}=\dot{\vec{r}}\times\vec{P}+\vec{r}\times\dot{\vec{P}}.
\label{Ldotrp}
\end{equation}
As for the velocity $\vec{v}=\dot{\vec{r}}$, we have from (\ref{vvprime})
\begin{equation}
\dot{\vec{r}}=\frac{\vec{P}}{m}+\vec{\Omega}\times\vec{r}.
\label{rdot}
\end{equation}
While, up to the first order terms in $\vec{H}$ and $\vec{\Omega}$,
\begin{equation}
\dot{\vec{P}}=\frac{d}{dt}\left (m\,\vec{v}-m\,\vec{\Omega}\times\vec{r}
\right )\approx -\frac{\alpha}{r^2}\,\vec{n}+\frac{e}{mc}\,\vec{P}\times
\vec{H}-\vec{\Omega}\times\vec{P}-m\dot{\vec{\Omega}}\times\vec{r}.
\label{pdot}
\end{equation}
Substituting (\ref{rdot}) and (\ref{pdot}) in (\ref{Ldotrp}), we get
$$\dot{\vec{L}}=(\vec{\Omega}\times\vec{r})\times\vec{P}+\frac{e}{mc}\,
\vec{r}\times(\vec{P}\times\vec{H})-\vec{r}\times(\vec{\Omega}\times\vec{P})
-m\vec{r}\times(\dot{\vec{\Omega}}\times\vec{r}),$$
or, after using $(\vec{\Omega}\times\vec{r})\times\vec{P}=-\vec{r}\times
(\vec{\Omega}\times\vec{P})-\vec{\Omega}\times(\vec{P}\times\vec{r})$,
$$\dot{\vec{L}}=\vec{\Omega}\times\vec{L}-\vec{r}\times\left [\left (
2\vec{\Omega}+\frac{e}{mc}\,\vec{H}\right)\times\vec{P}\right ]-
m\vec{r}\times(\dot{\vec{\Omega}}\times\vec{r}).$$
As we see, to get (\ref{Ldot}), it is sufficient to take
\begin{equation}
\vec{\Omega}=-\frac{e}{2mc}\,\vec{H}.
\label{OmegaH}
\end{equation}
Now let us calculate (note that $\dot{r}=\vec{v}\cdot\vec{n}$)
$$\dot{\vec{n}}=\frac{\vec{v}}{r}-\frac{\vec{r}(\vec{v}\cdot\vec{n})}{r^2}=
\frac{1}{r}\,(\vec{n}\times\vec{v})\times\vec{n}.$$
Substituting
$$\vec{v}=\frac{\vec{P}}{m}+\vec{\Omega}\times\vec{r}$$
and using
$$(\vec{n}\times\vec{P})\times\vec{n}=\frac{L}{r}\,\vec{e}_\varphi, \;\;\;
[\vec{n}\times(\vec{\Omega}\times\vec{n})]\times\vec{n}=
\vec{\Omega}\times\vec{n},$$
we get
\begin{equation}
\dot{\vec{n}}=\vec{\Omega}\times\vec{n}+\frac{L}{mr^2}\,\vec{e}_\varphi.
\label{ndot}
\end{equation}
Analogously,
$$\dot{\vec{e}}_\varphi=\dot{\vec{l}}\times\vec{n}+\vec{l}\times\dot
{\vec{n}}=(\vec{\Omega}\times\vec{l})\times\vec{n}+\vec{l}\times
(\vec{\Omega}\times\vec{n})+\frac{L}{mr^2}\,\vec{l}\times\vec{e}_\varphi.$$
But  $(\vec{\Omega}\times\vec{l})\times\vec{n}+\vec{l}\times(\vec{\Omega}
\times\vec{n})=-\vec{\Omega}\times(\vec{n}\times\vec{l})=\vec{\Omega}\times
\vec{e}_\varphi$ and $\vec{l}\times\vec{e}_\varphi=-\vec{n}$. Therefore,
\begin{equation}
\dot{\vec{e}}_\varphi=\vec{\Omega}\times\vec{e}_\varphi-\frac{L}{mr^2}\,
\vec{n}.
\label{e_phidot}
\end{equation}
At last, by using (\ref{larmor}) for $\dot{\vec{v}}$, we get
$$\dot{\vec{v}}^{\,\prime}=-\frac{\alpha}{mr^2}\,\vec{n}+\frac{e}{mc}\,
\vec{v}\times\vec{H}-\vec{\Omega}\times\vec{v}=\vec{\Omega}\times\vec{v}-
\frac{\alpha}{mr^2}\,\vec{n}.$$
But, up to first order terms in $\vec{\Omega}$, $\vec{\Omega}\times\vec{v}
\approx\vec{\Omega}\times\vec{v}^{\,\prime}$. Therefore,
\begin{equation}
\dot{\vec{v}}^{\,\prime}\approx \vec{\Omega}\times\vec{v}^{\,\prime}-
\frac{\alpha}{mr^2}\,\vec{n}.
\label{vprimedot}
\end{equation}
Having (\ref{vprimedot}) and (\ref{e_phidot}) at hand, it is easy to find
$$\dot{\vec{u}}\approx \vec{\Omega}\times\vec{v}^{\,\prime}-\frac{\alpha}
{L}\,\vec{\Omega}\times\vec{e}_\varphi=\vec{\Omega}\times\vec{u}.$$
Therefore, the Hamilton vector precesses with the same angular velocity
(\ref{OmegaH}) as the angular momentum, and this angular velocity can be
considered as related to the precession of the instantaneous ellipse as the
whole.

As the final remark, note that
$\vec{\cal{A}}=\frac{1}{2}\vec{H}\times\vec{r}$ 
is the vector-potential corresponding to the uniform magnetic field $\vec{H}$.
Therefore
$$\vec{P}=m\vec{v}-m\,\vec{\Omega}\times\vec{r}=m\vec{v}+\frac{e}{c}\,
\vec{\cal{A}}$$
is just the canonical momentum of the non-relativistic electron in the
magnetic field $\vec{H}$.

\section{Lense-Thirring precession}
It is well known \cite{4-1} that the vector-potential created by a magnetic
moment $\vec{\mu}$ is
\begin{equation}
\vec{\cal{A}}^{\,(\mu)}=\frac{\vec{\mu}\times\vec{r}}{r^3}.
\label{muA}
\end{equation}
Therefore, by making the change
$\vec{\mu}\to -\frac{2G}{c}\,\vec{L}^{\,\prime}$
in (\ref{muA}), we get the gravitational analog of the vector potential
for the gravitomagnetic field (\ref{egmf}) of the Earth with angular momentum
$\vec{L}^\prime$:
\begin{equation}
{\cal{A}}=\frac{2G}{cr^3}\,\vec{r}\times\vec{L}^{\,\prime}.
\label{gravA}
\end{equation}
To recast the canonical momentum,
$$\vec{P}=m\,\vec{v}+\frac{m}{c}\,{\cal{A}},$$
in the form $\vec{P}=m\,\vec{v}-m\,\vec{\Omega}\times\vec{r}$,
which corresponds to the decomposition of velocity (\ref{vvprime}), we could
take
\begin{equation}
\vec{\Omega}=\frac{2G}{c^2r^3}\,\vec{L}^{\,\prime}.
\label{OmegaG}
\end{equation}
Then the gravitomagnetic field (\ref{egmf}) can be rewritten as
\begin{equation}
\frac{\vec{H}}{c}=\vec{\Omega}-3\vec{n}\,(\vec{\Omega}\cdot\vec{n}).
\label{HcOmega}
\end{equation}
By using (\ref{HcOmega}) and equation of motion (\ref{seqm}), we get
\begin{equation}
\dot{\vec{P}}=-\frac{\alpha}{r^2}\,\vec{n}+2m\,\vec{v}\times\vec{\Omega}-
3m\,(\vec{\Omega}\cdot\vec{n})\,\vec{v}\times\vec{n}-m\,\dot{\vec{\Omega}}
\times\vec{r}.
\label{eqpd}
\end{equation}
From (\ref{OmegaG}), we find
$$\dot{\vec{\Omega}}=-3\frac{\vec{\Omega}}{r}\,(\vec{v}\cdot\vec{n}),$$
and (\ref{eqpd}) takes the form
$$\dot{\vec{P}}=-\frac{\alpha}{r^2}\,\vec{n}+2m\,\vec{v}\times\vec{\Omega}-
3m\,[\vec{v}\,(\vec{\Omega}\cdot\vec{n})-\vec{\Omega}\,(\vec{v}\cdot\vec{n})]
\times\vec{n}.$$
However,
$$[\vec{v}\,(\vec{\Omega}\cdot\vec{n})-\vec{\Omega}\,(\vec{v}\cdot\vec{n})]
\times\vec{n}=[\vec{n}\times(\vec{v}\times\vec{\Omega})]\times\vec{n}=
\vec{v}\times\vec{\Omega}-\vec{n}\,\,\vec{n}\cdot(\vec{v}
\times{\vec{\Omega}}),$$
and we end up with the equation
\begin{equation}
\dot{\vec{P}}=-\frac{\alpha}{r^2}\,\vec{n}+m\,\vec{\Omega}\times\vec{v}+
3m\,\vec{n}\,\,\vec{n}\cdot (\vec{v}\times\vec{\Omega}).
\label{eqpdf}
\end{equation}
Up to the first order in the small parameter $\vec{\Omega}$, (\ref{eqpdf})
can be rewritten as
\begin{equation}
\dot{\vec{P}}\approx -\frac{\alpha}{r^2}\,\vec{n}+\vec{\Omega}\times\vec{P}+
3\,\vec{n}\,\,\vec{\Omega}\cdot(\vec{n}\times\vec{P})
\label{pdotG}
\end{equation}
Now it is easy to find
\begin{equation}
\hspace*{-3mm}
\dot{\vec{L}}=\left (\frac{\vec{P}}{m}-\frac{1}{c}\,{\cal{A}}\right )\times
\vec{P}+\vec{r}\times(\vec{\Omega}\times\vec{P})=-(\vec{r}\times\vec{\Omega})
\times\vec{P}+\vec{r}\times(\vec{\Omega}\times\vec{P}).
\label{eqLd}
\end{equation}
But $\vec{r}\times(\vec{\Omega}\times\vec{P})+\vec{P}\times(\vec{r}\times
\vec{\Omega})=-\vec{\Omega}\times(\vec{P}\times\vec{r})$ and (\ref{eqLd})
indicates that the canonical angular momentum vector $\vec{L}$ really
precesses with angular velocity $\vec{\Omega}$:
\begin{equation}
\dot{\vec{L}}=\vec{\Omega}\times\vec{L}.
\label{LdotG}
\end{equation}
We are interested in the secular changes of the orbital parameters. Therefore
it makes sense to average $\vec{\Omega}$ in (\ref{LdotG}) over fast orbital
motion:
\begin{equation}
\vec{\Omega}\to \,<\vec{\Omega}>\,=\frac{2G}{c^2}\,\vec{L}^{\,\prime}\,
<\frac{1}{r^3}>.
\label{Omegaev}
\end{equation}
For the desired accuracy, we can average $1/r^3$ in (\ref{Omegaev}) over the
unperturbed orbit
$$\frac{p}{r}=1+e\cos{\varphi},$$
where $p$ and $e$ are the semi-latus rectum and eccentricity of the orbit.
But for the unperturbed orbit
\begin{equation}
dt=\frac{mr^2}{L}\,d\varphi,
\label{dtdphi}
\end{equation}
and we get
$$<\frac{1}{r^3}>=\frac{1}{T}\int\limits_0^T\frac{dt}{r^3}=\frac{m}{LTp}\int
\limits_0^{2\pi}(1+e\cos{\varphi})\,d\varphi=\frac{2\pi m}{LTp}.$$
Integrating (\ref{dtdphi}) over the complete orbital period $T$, we get
$$\frac{LT}{m}=2S=2\pi a^2\sqrt{1-e^2},$$
where $S$ is the area of the ellipse and $a$ is its semi-major axis. On the
other hand, $p=a(1-e^2)$ and we finally get
\begin{equation}
<\frac{1}{r^3}>=\frac{1}{a^3(1-e^2)^{3/2}}.
\label{ave1r3}
\end{equation}
Therefore, the averaged angular velocity of the precession is
\begin{equation}
<\vec{\Omega}>=\frac{2G}{c^2a^3(1-e^2)^{3/2}}\,\vec{L}^{\,\prime}.
\label{aveOmega}
\end{equation}

It is not difficult to check that relations (\ref{ndot}) and (\ref{e_phidot})
remain valid if $\vec{\Omega}$ in these formulas is given by (\ref{OmegaG}).
Then, after using (\ref{pdotG}) and (\ref{e_phidot}), we easily find
\begin{equation}
\dot{\vec{u}}\approx \vec{\Omega}\times\vec{u}+3\,\frac{\vec{n}}{mr}\,
\vec{\Omega}\cdot\vec{L}=\vec{\Omega}\times\vec{u}+3\,\frac{2G}{mc^2}\,
\frac{\vec{n}}{r^4}\,\vec{L}^{\,\prime}\cdot\vec{L}.
\label{equdotG}
\end{equation}
Therefore, the Hamilton vector does not precess with angular velocity
$\vec{\Omega}$. However, we should average (\ref{equdotG}) over the fast
orbital motion. The first term simply gives $<\vec{\Omega}\times\vec{u}>\,=\,
<\vec{\Omega}>\times\;\vec{u}$ because $\vec{u}$ is a slowly changing vector
and we can assume that it does not change over time scales comparable to the
orbital period $T$. As for the second term, we can use $\vec{n}=\cos{\varphi}
\,\vec{i}+\sin{\varphi}\,\vec{j}$ and get
\begin{equation}
<\frac{\vec{n}}{r^4}>=\frac{m}{TLp^2}\int\limits_0^{2\pi} (\cos{\varphi}\,
\vec{i}+\sin{\varphi}\,\vec{j})(1+e\cos{\varphi})^2\,d\varphi=\frac{2\pi me}
{TLp^2}\,\vec{i}.
\label{eqnr4}
\end{equation}
However,
$$\vec{i}=\vec{j}\times\vec{k}=\frac{\vec{u}}{u}\times\frac{\vec{L}}{L},
\;\;\; \frac{e}{puL}=\frac{1}{p\alpha}=\frac{m}{L^2},$$
and (\ref{eqnr4}) takes the form
\begin{equation}
<\frac{\vec{n}}{r^4}>=<\frac{1}{r^3}>\frac{m}{L^2}\,\vec{u}\times\vec{L}.
\label{avenr4}
\end{equation}
Therefore, after averaging, (\ref{equdotG}) changes to
\begin{equation}
\dot{\vec{u}}\approx \vec{\Omega}_{LT}\times\vec{u},
\label{udotG}
\end{equation}
where
\begin{equation}
\vec{\Omega}_{LT}=\frac{2G}{c^2a^3(1-e^2)^{3/2}}[\vec{L}^{\,\prime}-3\vec{l}
\,(\vec{l}\cdot \vec{L}^{\,\prime})].
\label{OmegaLT}
\end{equation}
As we see, the secular precession of the elliptic orbit contains two terms:
the precession of the orbital plane with the angular velocity $<\vec{\Omega}>$
around the central body's angular momentum $\vec{L}^{\,\prime}$, and the
precession within the orbital plane with the angular velocity $\vec{\Omega}
_{LT}\;-<\vec{\Omega}>$ around the angular momentum $\vec{L}$. The magnitudes
of the  angular momentum $\vec{L}$ and the Hamilton vector $\vec{u}$, and
therefore, the orbital parameters such as the eccentricity and the semi-major
axis, remain unchanged to the first order of the perturbation theory.

\section{Concluding remarks, Skovoroda's principle and all that}
Lev Borisovich Okun cites \cite{5-1} eighteenth century Ukrainian
philosopher Grigory Skovoroda as the author of the remarkable
principle, which Okun's PhD adviser Isaak Yakovlevich Pomeranchuk
used to quote: ``Thanks God: All what is relevant is simple, all
what is not simple is not relevant.'' The formal analogy between
weak field low velocity general relativity and Maxwellian
electrodynamics is a simple and elegant way to illuminate a whole
class of interesting physical phenomena dubbed gravitomagnetism.
Lense-Thirring precession is one such example. However, it should
be kept in mind that the analogy is only formal and sometimes can
lead to strange and erroneous conclusions if we forgot about
rather strong limitations under which the approximation underlying
the analogy is valid \cite{5-2}.

Frame dragging is maybe more appropriate interpretation of the Lense-Thirring
effect not restricted to the weak field limit. However, ``whereas `frame
dragging' is a very catchy appellation'' \cite{5-3} its meaning is not easy
to explain to introductory level students. Therefore, for them and not only
explaining the Lense-Thirring effect by analogy with Larmor precession remains
a relevant and simple option.

The use of Hamilton or Runge-Lenz vector greatly simplifies the discussion.
However, a great deal of vector algebra is still needed in either cases, as
the previous chapters illustrate. This is in contrast with situation in
perihelion precession under central force perturbations where the use of the
Hamilton vector trivializes the problem \cite{5-4}. In fact, the amount of
vector algebra is a price we should pay for our desire to keep the approach
elementary and accessible in introductory mechanics course. If more advanced
background in analytical mechanics is assumed, much more elegant and
technically simple way is provided by the use of the Poisson brackets
\cite{5-5} or, alternatively, by Hamilton equations of motion \cite{5-6}.

In terms of the standard Poisson brackets, we have
\begin{equation}
\dot{L}_i=\{L_i, {\cal{H}}\},\;\;\;\;
\dot{u_i}=\{u_i, {\cal{H}}\},
\label{LuPB}
\end{equation}
where the Hamiltonian has the form ${\cal{H}}={\cal{H}}_0+\delta{\cal{H}}$
with $\delta{\cal{H}}=\vec{\Omega}\cdot\vec{L}$ \cite{5-6} and
$\{L_i, {\cal{H}}_0\}=\{u_i, {\cal{H}}_0\}=0$.

The Poisson brackets are easy to calculate by using the Leibniz rule
$\{f,gh\}=\{f,g\}h+g\{f,h\}$ and the fundamental Poisson brackets
\begin{eqnarray} &
\{r_i, r_j\}=0,\;\;\; & \{P_i, P_j\}=0, \hspace*{12mm} \{r_i, P_j\}=
\delta_{ij}, \nonumber \\ &
\{L_i, r_j\}=\epsilon_{ijk}\,r_k,\;\;\;& \{L_i, P_j\}=\epsilon_{ijk}\,P_k,
\;\;\; \{L_i, L_j\}=\epsilon_{ijk}\,L_k,\nonumber \\ & \{L_i, f(\vec{r},
\vec{p})\}=0, \;\;\;& \{P_i, f(\vec{r},\vec{p})\}=-\frac{\partial f(\vec{r},
\vec{p})}{\partial r_i},
\label{fPB}
\end{eqnarray}
where $f(\vec{r},\vec{p})$ is any scalar function of its arguments. It follows
from (\ref{fPB}) that
\begin{eqnarray} &
\{L_i, \Omega_j\}=0, \;\;\;&  \{L_i, e_{\varphi j}\}=\epsilon_{ijk}\,
e_{\varphi k}, \nonumber \\ &
\{\Omega_i, e_{\varphi j}\}=0,\;\;\;& \{P_i, \Omega_j\}=3\,\frac{\Omega_j}{r}
\,n_i.
\label{iPB}
\end{eqnarray}
Therefore,
$$\dot{L}_i=\{L_i, \Omega_j\}\,L_j+\{L_i, L_j\}\,\Omega_j=\epsilon_{ijk}\,
L_k\Omega_j,$$
which is equivalent to (\ref{LdotG}).

Besides, for any vector of the form $\vec{B}=f_1(\vec{r},\vec{p})\,\vec{r}+
f_2(\vec{r},\vec{p})\,\vec{P}$, with any scalar functions $f_1$ and $f_2$, and
in particular for vectors $\vec{e}_\varphi$ and $\vec{u}$, we will have
$$\{L_i, B_j\}=\epsilon_{ijk}\,B_k,$$
and we get immediately
$$\dot{u}_i=\{u_i, L_j\}\Omega_j+\frac{1}{m}\{P_i, \Omega_j\}L_j=
\epsilon_{ijk}\,u_k\Omega_j+\frac{3n_i}{mr}\,\vec{\Omega}\cdot\vec{L},$$
which is equivalent to (\ref{equdotG}).

As we see, in combination with some background in Hamiltonian mechanics, the
use of Hamilton vector again makes the exposition rather trivial, in complete
agreement with the Skovoroda's principle.

In our discussions canonical variables played an important role, as they are
in accord with the intuitive physical picture of precessing orbital plane.
However, the use of canonical variables are not mandatory for the perturbation
theory discussion of the Lense-Thirring effect. The standard perturbation
theory in celestial mechanics is based on variation-of-parameters method
which has some inherent gauge freedom \cite{5-7}. Indeed, the unperturbed
Keplerian ellipse is determined by six parameters
\begin{equation}
\vec{r}=\vec{f}(C_1,C_2,C_3,C_4,C_5,C_6,t).
\label{upell}
\end{equation}
In the role of these parameters one can take, for example, three Euler angles
which determine the orbit plane orientation, the semi-major axis and
eccentricity of the orbit, and the so called mean anomaly at an epoch, which
determines the initial position of the body. Another obvious choice is the
initial values of the position and velocity vectors of the body, and many
other sets (Delaunay, Poincare, Jacobi, Hill) can be found in the literature
\cite{5-8}.

To solve the perturbed equation
\begin{equation}
m\ddot{\vec{r}}=-\frac{\alpha}{r^2}\,\vec{n}+\vec{F},
\label{pereqm}
\end{equation}
with $\vec{F}$ as a small perturbation, by the variation-of-parameters method,
one can assume that the solution still has the form (\ref{upell}) but $C_i$
are no longer constant. However, three scalar equations (\ref{pereqm}) are not
sufficient to determine six unknown functions $C_i(t)$. We need three
auxiliary conditions on them and the freedom in choosing of these auxiliary
conditions is just the gauge freedom mentioned above. Usually the gauge fixing
is achieved by the Lagrange constraint
\begin{equation}
\sum\limits_{i=1}^6 \dot{C}_i\,\frac{\partial \vec{f}}{\partial C_i}=0.
\label{LagrangeC}
\end{equation}
Under this condition
$$\dot{\vec{r}}=\frac{\partial \vec{f}}{\partial t}.$$
That is, the velocity is the same function of the parameters $C_i$ as in the
absence of perturbations and the instantaneous ellipse, determined by these
parameters, is tangent to the real trajectory. Correspondingly, the
instantaneous orbital parameters $C_i$ so determined are called osculating.

However, not all choices of orbital parameters $C_i$ are compatible to the
Lagrange constraint (\ref{LagrangeC}). For example, in previous chapters we
have choose the canonical angular momentum and the Hamilton vector to
characterize the orbit shape and orientation. When velocity dependent
perturbations are present, such parameters are not osculating, as is evident
from the ascribed physical picture of the combined motion.

The canonical perturbation theory in celestial mechanics was developed in
\cite{1-7}. Again the corresponding orbital elements are not osculating if
the perturbation depends on velocity. In fact this is a general property:
under velocity-dependent disturbances, canonicity and osculation are not 
compatible \cite{5-7}.

The problem of orbit perturbations can be solved in any gauge. At that some
gauges are more convenient, because if the gauge corresponds to the real
physical picture of the perturbed motion, the formalism simplifies. Some
examples are given in \cite{1-7} when the standard perturbation theory with
osculating elements is cumbersome while the canonical perturbation theory
is more elegant.

The Lense-Thirring effect can be considered either in the canonical
perturbation theory, or with the osculating orbital elements \cite{1-7},
or in any other convenient gauge. At that the corresponding orbital elements
can differ considerably. Of course, the real physical quantities, like
position and velocity, do not depend on the gauge used after the initial
values are accounted correctly \cite{1-7}. However, sometimes the orbital
elements, like inclination angle, are also considered as physically real.
In such cases care should be taken to relate the measured quantities to the
orbital parameters used in the perturbation theory. Without this care a
confusion can arise when two different mathematically correct approaches
give seemingly different results which are in fact equivalent \cite{1-6}.

\section*{Acknowledgments}
The work of Z.K.S. is supported in part by grants Sci.School-905.2006.2 and
RFBR 06-02-16192-a.

\end{document}